\pdfoutput=1

\documentclass[
  twocolumn,
  prl,
  showpacs,
  amsmath,
  amssymb,
  superscriptaddress
]{revtex4}

\usepackage{bm}
\usepackage{graphicx}
\usepackage{amssymb}
\usepackage{xcolor}

\newcommand{\tr}{\text{tr}}
\newcommand{\Tr}{\text{Tr}}
\renewcommand{\v}[1]{\textbf{\textit #1}}
\newcommand{\sgn}[1]{\text{sgn}({#1})}

\begin{document}

\title{Anomalous Hall Effect on the surface of topological Kondo insulators}

\author{E.\ J.\ K\"onig}
\affiliation{Department of Physics, University of Wisconsin-Madison, Madison, Wisconsin 53706, USA}

\author{P.\ M.\ Ostrovsky}
\affiliation{Max Planck Institute for Solid State Research, Heisenbergstr. 1, 70569 Stuttgart, Germany}
\affiliation{L.\ D.\ Landau Institute for Theoretical Physics RAS, 119334 Moscow, Russia}

\author{M.\ Dzero}
\affiliation{Department of Physics, Kent State University, Kent, OH, 44242, USA}

\author{A.\  Levchenko}
\affiliation{Department of Physics, University of Wisconsin-Madison, Madison, Wisconsin 53706, USA}
\affiliation{Department of Physics and Astronomy, Michigan State University, East Lansing, Michigan 48824, USA}

\begin{abstract}
We calculate the anomalous Hall conductivity $\sigma_{xy}$ of the surface states {in cubic topological Kondo insulators}. We consider a generic model for the surface states with three Dirac cones on the (001) surface. The Fermi velocity, the Fermi momentum and the Zeeman energy in different Dirac pockets may be unequal. The microscopic impurity potential mediates mixed intra and interband extrinsic scattering processes. Our calculation of $\sigma_{xy}$ is based on the Kubo-Streda diagrammatic approach. It includes diffractive skew scattering contributions originating from the rare two-impurity complexes. Remarkably, these contributions yield anomalous Hall conductivity that is independent of impurity concentration, and thus is of the same order as other known extrinsic side jump and skew scattering terms. We discuss various special cases of our results and the experimental relevance of our study in the context of the recent hysteretic magnetotransport data in SmB$_6$ samples.
\end{abstract}

\date{January 2, 2016} 


\maketitle

\paragraph{\textbf{Topological Kondo Insulators.}} Topological insulators~\cite{HasanKane2010, QiZhang2011, Bernevig2013} remain a vibrant field {of research} in present day condensed matter physics. The main thrusts for this extraordinary scientific interest include the vast potential technological applications in the fields of nanoelectronics and quantum computation {as well as}  the fascinating innovative realization of fundamental concepts from quantum field theory and differential geometry. 

%%%%%%%%%%%%%%%%%%%%%%%%%%%%%%
%%%%%%%%%%%%%%%%%%%%%%%%%%%%%%
\begin{figure}[t]
\includegraphics[scale=.45]{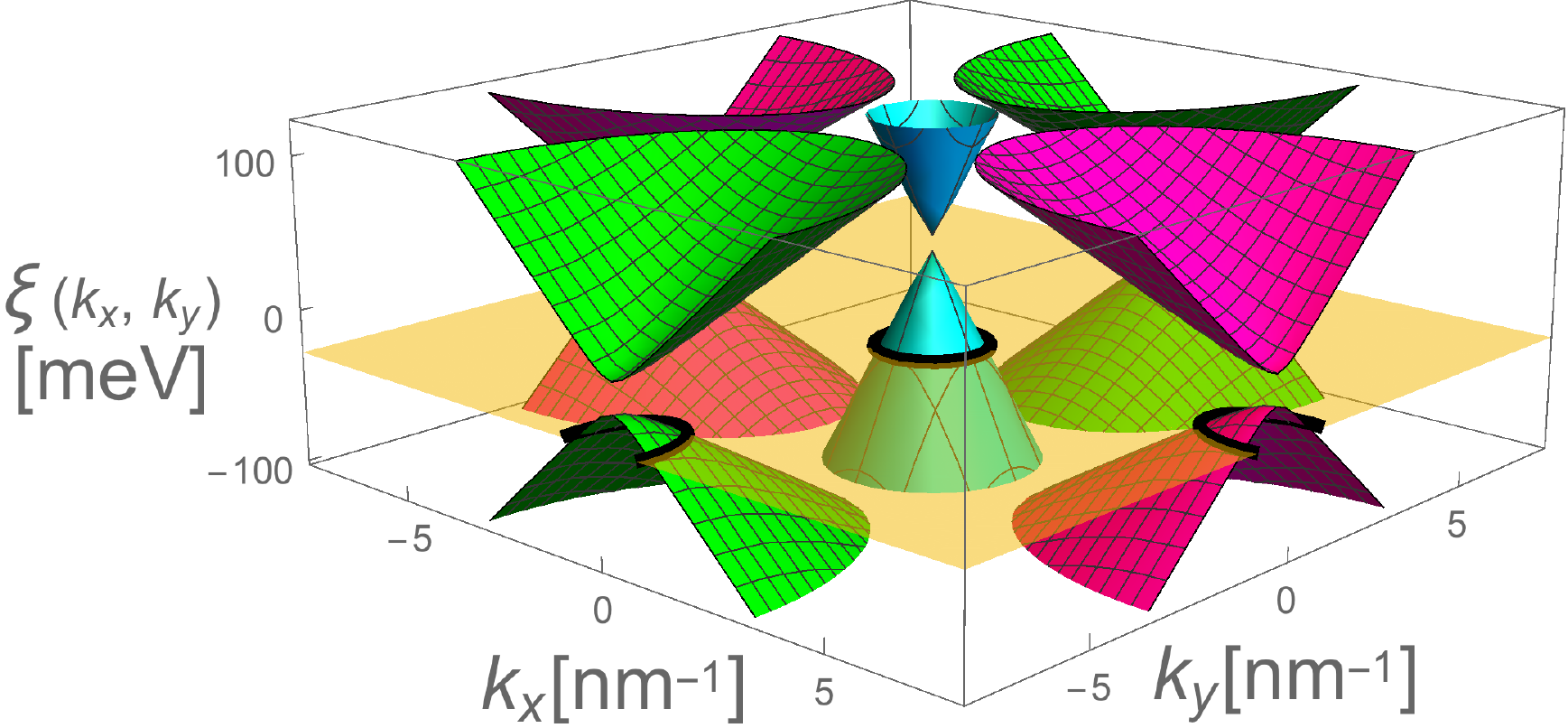} 
\caption{Dispersion relation of surface states in a TKI with cubic symmetry. We chose experimentally  realistic parameters for the case of SmB$_6$ \cite{NeupaneHasan2013}: Fermi velocities of $\simeq$ 72 (26) $\textrm{meV}\cdot\textrm{nm}$ at the $\Gamma$ ($X$ and $Y$) points, an offset $E_\Gamma \approx 39 \, \textrm{meV}$ of the central Dirac cone, an ellipticity $\sqrt{v_{X,y}/v_{X,x}} \approx 1.2$, a reciprocal lattice constant $2 \pi/a = 15~\textrm{nm}^{-1}$. Further, we assumed a gap of $10 \, \textrm{meV}$ ($20 \, \textrm{meV}$) at the $\Gamma$ ($X$ and $Y$) points.}
\label{fig:DispRelSmB6}
\end{figure}
%%%%%%%%%%%%%%%%%%%%%%%%%%%%%%
%%%%%%%%%%%%%%%%%%%%%%%%%%%%%%

Among the various realizations of topological phases of matter, topological Kondo insulators (TKIs)~\cite{DzeroColeman2010,DzeroColeman2015}, take a special place. Their topologically protected metallic surface states emerge as a result of the hybridization between weakly correlated conduction electrons and strongly correlated states. In particular, theories \cite{LuDai2013,AlexandrovColeman2013,YeSun2013,RoyGalitski2014} describing states on the (001) surface suggest a low energy Hamiltonian with three Dirac bands located at $\Gamma$, $X$ and $Y$ points of the surface Brillouin zone (BZ), see Fig.~\ref{fig:DispRelSmB6}.  Main experimentally distinguishing characteristics of the TKIs are the saturation of resistivity at very low temperatures, pronounced temperature dependence of the magnetic susceptibility across a wide range of temperatures and a fairly narrow insulating gap \cite{MenthBuehlerGeballe1969,{AllenWachter1979}}. 
Intriguing recent experimental evidence for the TKI physics was reported in SmB$_6$ samples revealing the predicted surface dominated transport directly~\cite{WolgastFisk2013}, by thickness independent resistivity measurements showing the violation of Ohm's law~\cite{KimXia2013}, and 2D (surface) weak antilocalization data~\cite{ThomasXia2013}. Furthermore, characteristics of Dirac electrons were revealed using ARPES \cite{NeupaneHasan2013,XuShi2013,JiangFeng2013} and torque magnetometry \cite{LiLi2014}. In addition, hysteretic magneto-transport measurements have been reported by several groups~\cite{EoFisk2014, NakajimaPaglione2013}. This effect can be attributed to ferromagnetic domains formed on the surface by unscreened samarium magnetic moments or samarium sesquioxide (Sm$_2$O$_3$) impurities.

It has been also proposed that the surface states in SmB$_6$ may be of conventional type~\cite{ZhuDamiscelli2013, HlawenkaRienks2015}, i.e. they have quadratic dispersion modified by the presence of strong spin-orbit coupling. Note, that in this scenario, the conduction states will remain decoupled from the Sm moments on the surface via the Kondo breakdown mechanism, so that ferromagnetic ordering of the samarium $f$-electrons would still be possible. This controversy -- 
Dirac vs. conventional surface states -- motivates us to study the magnetotransport properties of the surface states on the background of induced nonzero magnetization. Specifically, in this paper we calculate the anomalous Hall conductivity for a cubic topological Kondo insulator with three Dirac surface bands. 
Our results for the anomalous Hall conductivity allow us to elucidate experimentally distinguishable characteristics of the Dirac electrons and should help 
to resolve the controversy discussed above. 

%%%%%%%%%%%%%%%%%%%%%%%%%%%%%%%%
%%%%%%%%%%%%%%%%%%%%%%%%%%%%%%%%
\begin{figure}
\includegraphics[scale=.45]{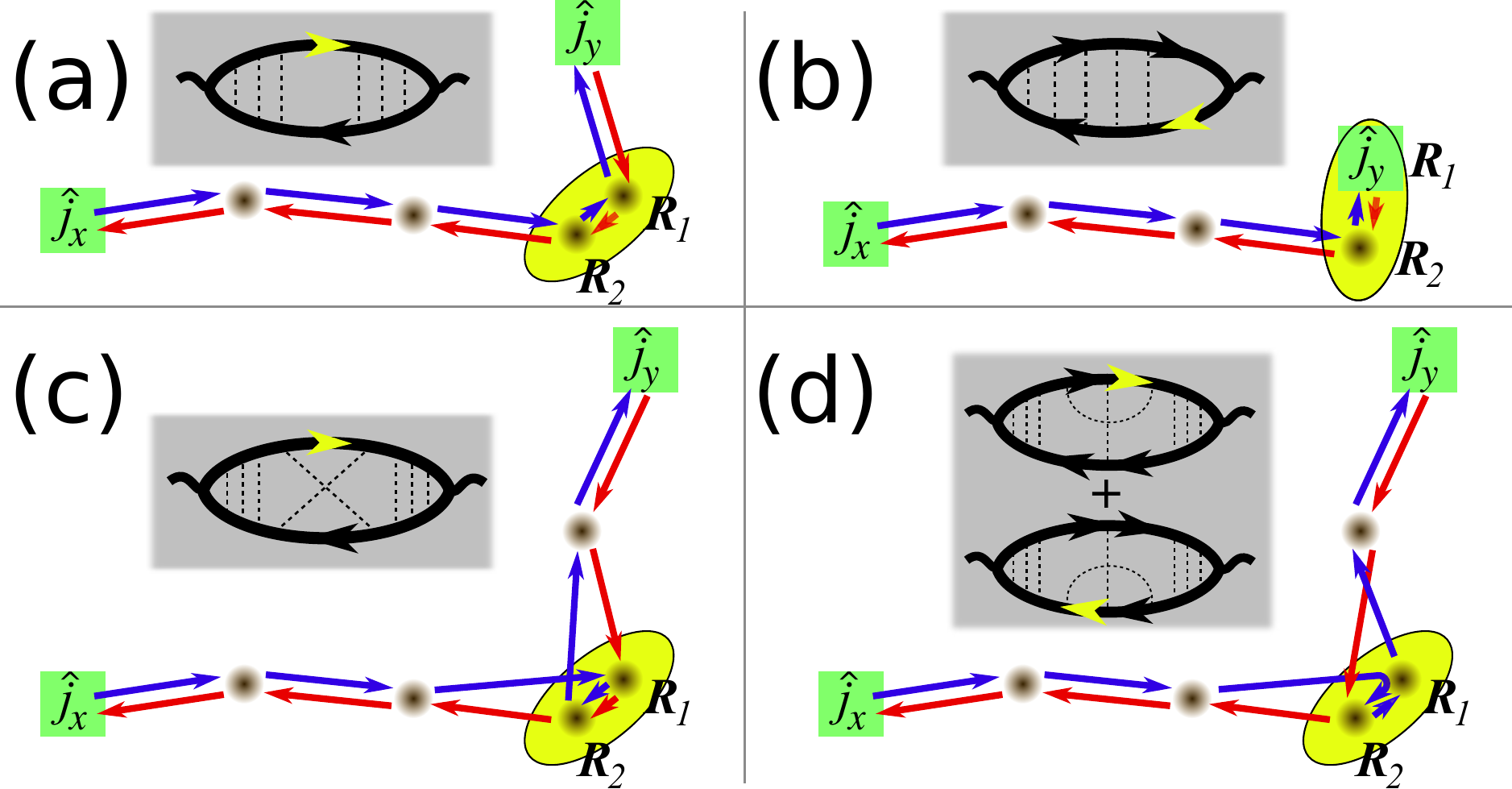} 
\caption{Diagrammatic representation and real space trajectories for extrinsic contributions to $\sigma_{xy}$. Quantum complexes responsible for the AHE are shown by an ellipse with focuses in points $\bm{R_1}$ and $\bm{R_2}$. In the noncrossing approximation, both skew-scattering (a) and side jump (b) contributions rely on coherent interband scattering between opposite branches of the Dirac spectrum. The corresponding virtual states as well as off-shell excitations entering crossed X and $\Psi$ diagrams, (c) and (d) respectively, are marked by a yellow arrow in exemplary positions of the diagrams. Due to the uncertainty principle, the typical extension of a quantum complex is thus of the order of Fermi wavelength $|\bm{R_1}-\bm{R_2}|\sim\lambda_F$.}
\label{fig:Condbubbles}
\end{figure}
%%%%%%%%%%%%%%%%%%%%%%%%%%%%%%%%
%%%%%%%%%%%%%%%%%%%%%%%%%%%%%%%%

\paragraph{\textbf{Anomalous Hall effect.}} Electron transport in ferromagnets has a long history going back to E. Hall's 1881 discovery of the anomalous Hall effect (AHE)~\cite{NagaosaOng2010}, i.e.~of a transverse conductivity $\sigma_{xy}$ generated by the magnetization (Zeeman coupling) rather than by orbital coupling to a magnetic field. To account for this effect, two equally appropriate techniques are commonly employed. First, in the semiclassical approach~\cite{Sinitsyn2008} different terms in $\sigma_{xy}$ stem from distinct physical mechanisms of intrinsic~\cite{KarplusLuttinger1954}, skew scattering~\cite{Smit1955} and side jump~\cite{Berger1964} contributions. Second, $\sigma_{xy}$ can be directly calculated using Kubo-Streda diagrammatic response theory~\cite{Streda}. Semiclassics appear to be more intuitive, while the diagrammatic treatment is more systematic. Notably, an additional skew scattering mechanism was discovered only very recently with the help of diagrams \cite{SinitsynSinova2007,AdoTitov2015,AdoTitov2016}. Physically, it originates from \textit{diffractive skew scattering} off two impurities residing about one Fermi wavelength $\lambda_F$ from each other. Diagrammatically, these processes can be understood by considering crossed impurity lines in the conductivity bubble [see Fig.~\ref{fig:Condbubbles} (c), (d)], and can be equivalently treated in the semiclassical approach provided that crossed impurity lines are included into the full scattering amplitude. At first glance, this observation seems to be in sharp contrast with conventional wisdom of the impurity diagrammatic technique \cite{AGD} that dictates that a single cross of impurity lines implies a rare disorder configuration and thus smallness in the parameter $\lambda_F/l\ll1$, where $l$ is the elastic mean free path. However, it should be stressed that even previously discussed \cite{Sinitsyn2008, KarplusLuttinger1954, Smit1955, Berger1964, SinitsynSinova2007, SinitsynMacDonald2006} contributions of weak impurities to the AHE rely on rare impurity configurations (see Fig.~\ref{fig:Condbubbles}). Therefore, both crossed and non-crossed diagrams are of the same order and suppressed by $\lambda_F/l$ as compared to the diagonal conductance. Diagrams with more than a single cross are even smaller \cite{FootnoteInterference}. These qualitative arguments are fully supported by the microscopic computation which we present in the remainder of the paper. 

\begin{figure}[t]
\includegraphics[scale=.45]{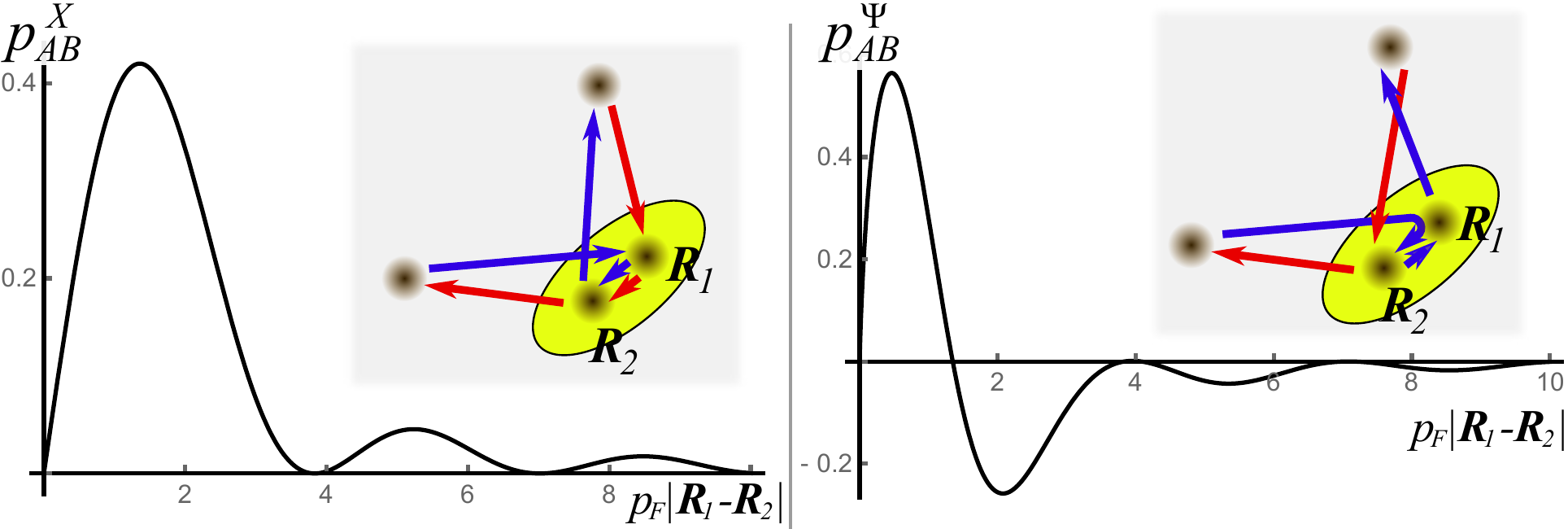} 
\caption{Diffractive skew scattering: Spatially-resolved scattering probabilities $p_{AB}^X$ and  $p_{AB}^\Psi$ for intraband scattering.}
\label{fig:Amplitudes}
\end{figure}

To make the diffractive analogy transparent we present in Fig.~\ref{fig:Condbubbles} examplary electron trajectories in real space. The probability $p_{AB}= \vert \sum_i A_i \vert^2 = \sum_{i j} A_i A_j^*$ of an electron reaching a point $\v r_B$ from $\v r_A$ is the square of the sum of the amplitudes for all paths $i,j$. In Fig.~\ref{fig:Condbubbles}, $A_i$ and $A_j^*$ are represented by different colors and opposite orientation of arrows. In the noncrossing approximation, $p_{AB}^{(nc)} = \sum_i \vert A_i\vert^2$, and all interference terms are omitted. The crossed $X$-diagrams contribute $p_{AB}^{X} = \sum_{i \neq j} ' A_i A_j^*$, where the sum includes pairs of nonequal trajectories equivalent to Fig.~\ref{fig:Condbubbles} (c). An analogous expression holds for $p_{AB}^\Psi$. The interference pattern becomes apparent in the plots of spatially-resolved scattering probabilities $p_{AB}^{X,\Psi}$ off two-impurity complexes, see Fig.~\ref{fig:Amplitudes}. The latter exhibit pronounced Fraunhofer oscillatory interference patterns. The novel extrinsic contributions [Fig. 2 (c,d)] constitute inherent parts of the skew scattering and should necessarily be included to properly compute the transverse conductivity. These terms can be distinguished from previously studied processes [Fig. 2 (a,b)] by means of their diffractive nature. 

%%%%%%%%%%%%%%%%%%%%%%%%%%%%%%
%%%%%%%%%%%%%%%%%%%%%%%%%%%%%%
%%%%%%%%%%%%%%%%%%%%%%%%%%%%%%
%%%%%%%%%%%%%%%%%%%%%%%%%%%%%%

\paragraph{\textbf{Model and Assumptions.}} We employ the diagrammatic approach to calculate the anomalous Hall response on the surface of 3D TKIs with cubic symmetry taking into account all the diagrams to the leading order in impurity concentration. For this purpose, consider the following low energy Hamiltonian
\begin{equation}
\underline{H}_0 =  \sum_{K } \left[v_K \boldsymbol{\sigma} \cdot \v p + m_K \sigma_z + E_K\right] \underline \Pi_K . \label{eq:H0}
\end{equation}
Throughout the paper, matrices in the space of Dirac pockets (DPs) are denoted by an underscore and have indices $K, K' \in \lbrace \Gamma,X,Y \rbrace$. The symbol $\underline \Pi_K$ denotes a projector on $K$th DP. Rotational $C_4$ symmetry imposes $v_X = v_Y \equiv v$, $m_X = m_Y \equiv m$. We count energies from the Dirac point of the $X$ pocket, $E_X = E_Y \equiv 0$, and momenta in $X$ ($Y$) pocket relative to ${\v Q}_{X (Y)} = (\pi/a)\hat{\v e}_{X (Y)} $ ($a$ is the lattice spacing). For simplicity we omit the anisotropy of $X$ and $Y$ pockets and set $\hbar = 1$ in the intermediate formulas (we restore Planck's constants in the final expressions for $\sigma_{xy}$). The Hamiltonian in Eq.~\eqref{eq:H0} contains just two essential ingredients for the finite AHE, namely spin-orbit coupling and magnetization (time reversal symmetry breaking), which is implicit in the mass term of the Dirac fermions.  

Our model also contains uniformly distributed scalar impurities with isotropic potential $u(\vert \v r \vert)$ which is short-ranged on the scale of the smallest Fermi wavelength $\min_K (p_{F,K}^{-1})$ with $v_K p_{F,K} = \sqrt{\epsilon_K^2 - m_K^2}$ where $\epsilon_K = \epsilon - E_K$ and $\epsilon$ is the Fermi energy. Our calculation is controlled in the parameter $n_{\rm imp}/ n_{\rm min}\ll 1$ with $n_{\rm imp}$  and $n_{\rm min} = \min_K (n_K)$ being the impurity concentration and the carrier density of the least populated pocket respectively. We assume weak impurities and treat them in the leading Born approximation.
 
Technically, the anomalous Hall response involves virtual states (off-shell contributions) residing within the radius $\Delta p= 3 \max_K (p_{F,K})$ around $\Gamma$, $X$, $Y$ and $M$ points of the BZ. As a consequence, the minimal three-band model Eq.~\eqref{eq:H0} is applicable only for sufficiently small Fermi momenta $ \max_K (p_{F,K}) \ll \pi/a$ (see yellow plane in Fig.~\ref{fig:DispRelSmB6}). Furthermore, the contribution to $\sigma_{xy}$ originating from the states in the vicinity of the $M$ point is negligible provided $\sqrt{2 m_M \Delta} \gg \min_K (p_{F,K})$, where $m_M$ ($\Delta$) is the effective mass (excitation gap of closest states) at the $M$ point~\cite{Supplmat}.

\paragraph{\textbf{Calculation and results.}}
It is common to distinguish the following two contributions to the anomalous Hall conductivity $\sigma_{xy}=\sigma^{I}_{xy}+\sigma^{II}_{xy}$:
\begin{equation}
\sigma_{xy}^I = \frac{e^2}{h} \left\langle \Tr \left[\underline{\hat j}_x \underline{G}^R \underline{\hat j}_y \underline{G}^A\right] \right\rangle, \quad
\sigma_{xy}^{II} = ec \sum_K \left . \frac{\partial n_K}{\partial B}\right \vert_{B = 0}. \label{eq:KuboStreda}
\end{equation}
The angular brackets denote disorder average in this expression. The bare current operators are $\underline{\hat j}_\mu = \text{diag}\left ( v_\Gamma\sigma_\mu, v \sigma_\mu, v \sigma_\mu \right )$, while the clean Green's functions at the Fermi energy $\epsilon$ are $\underline{G}^{R/A}_0 = (\epsilon \pm i 0- \underline{H}_0)^{-1}$.

The disorder average leads to a finite self-energy entering the Green's function. We find in momentum space, 
\begin{equation}
\underline{G}^R( \v p, \epsilon) = \sum_{K}  \frac{\epsilon_K^+ + v_K \boldsymbol \sigma\cdot\v p + m_K^- \sigma_z}{(\epsilon_K^+)^2 - [(v_Kp)^2 + (m_K^-)^2]} \underline{\Pi}_K\label{eq:GSCBA}
\end{equation}
with $\epsilon^{\pm}_K =  \epsilon_K \pm i \boldsymbol \Gamma_K$, $m^{\pm}_K = m_K \pm i \boldsymbol \Gamma_K^{(m)}$. Here we also introduced the total scattering rates in the pocket $K$
\begin{subequations}
\begin{eqnarray}
\hskip-.45cm
&&\boldsymbol \Gamma_K = \sum_{K' }  \Gamma_{K \rightarrow K' \rightarrow K}=\sum_{K'} \frac{\pi}{2}\nu_{K'}(\epsilon_{K'}) [\underline{ W}]_{K'K} \, , \\
\hskip-.45cm
&&\boldsymbol \Gamma_K^{(m)} = \sum_{K' }   \Gamma^{(m)}_{K \rightarrow K' \rightarrow K} =\sum_{K'}\Gamma_{K \rightarrow K' \rightarrow K}\frac{m_{K'}}{\epsilon_{K'}}\, ,
\end{eqnarray}
\label{eq:Gammas}
\end{subequations}
as a function of intra- ($K'=K$) and interpocket ($K'\not=K$) scattering rates $
\Gamma_{K \rightarrow K' \rightarrow K} $ and $\Gamma_{K \rightarrow K' \rightarrow K}^{(m)}$. Here, $\nu_K (\epsilon_K) =  \theta( \epsilon_K^2 - m_K^2)\vert \epsilon_K\vert/2\pi v_K^2$ is the density of states. Furthermore we introduced the matrix 
\begin{equation}
\underline{ W} = \left (\begin{array}{ccc}
 W &  W_{\Gamma X} &  W_{\Gamma X} \\ 
 W_{\Gamma X} &  W &  W_{XY} \\ 
 W_{\Gamma X} &  W_{XY} &  W
\end{array} \right )
\end{equation}
with entries $W= n_{\rm imp} \vert \tilde u(0) \vert^2$, $W_{\Gamma X}=n_{\rm imp} \vert \tilde u(\pi/a) \vert^2$, and $W_{X Y}=n_{\rm imp} \vert \tilde u(\sqrt{2}\pi/a) \vert^2$, where $\tilde u(\v q)$ is the Fourier transform of $u(\v r)$.

We now turn our attention to the Hall response. When the Fermi energy lies in the gap, $\vert \epsilon_K \vert < \vert m_K\vert$, the contribution of the $K$th DP to the Hall response is 
\begin{equation}
\left . \sigma_{xy} \right \vert_K  = -\frac{\sgn{m_K}}{2} \frac{e^2}{h}, \label{eq:sigmaxygap}
\end{equation}
and stems from $\sigma_{xy}^{II}$, only. The half-integer quantization is a consequence of fermion number fractionalization~\cite{Supplmat, JackiwRebbi1978, KoenigMirlin2014}. This result can be understood in terms of the intrinsic mechanism, so called \textit{anomalous velocity} contribution to AHE, as originally introduced by Karplus and Luttinger \cite{KarplusLuttinger1954}. Its topological origin was realized much later \cite{Chang-Niu-PRL95}, and can be equivalently understood in terms of the Berry curvature that acts as an effective magnetic field for electron wave-packet motion in parameter space of momenta. Indeed, Hall conductivity can be presented as  $\sigma_{xy}|_K=-(e^2v^2_K/2\pi h)\int\Omega_{xy}(k)d^2k$, where the Berry curvature for a single gapped Dirac cone is given explicitly by $\Omega_{xy}(k)=m_K/2(m^2_K+v^2_Kk^2)^{3/2}$ so that upon momentum integration Eq.~\eqref{eq:sigmaxygap} follows. 

In contrast, outside the gap the contribution of $\sigma_{xy}^{II}$ is subleading in $n_{\rm imp}/n_{\rm min}\ll1$. We therefore now focus on 
the contribution of $\sigma_{xy}^{I}$. We switch to a matrix representation in DP space and find 
\begin{equation}
\sigma_{xy}^{I} = 2\frac{e^2}{h} \underline{\v v} \underline F \left[ \underline{{b}} +\underline a [\underline{x} + \underline{ \psi}]\underline a\right] \underline F^T \underline{\v v}^T. \label{eq:sigmaxyIresult}
\end{equation}			
In this expression, the bare velocity vertex is $\underline{\v v} = (v_\Gamma, v, v)$ and the trace over spin space was already performed. The various contributions have the following origin [cf. Fig.~\ref{fig:Condbubbles}]: $\underline{F}$ is the noncrossed vertex correction, in which at each stringer $\underline a$ of the ladder only contributions which are on-shell were kept; $\underline{b}$ is also part of the ladder in diagram, but involves contributions away from the Fermi surface; finally $\underline{x}$ [$\underline{\psi}$] originates from the central part of diagrams (c) and (d). All the elements of Eq.~\eqref{eq:sigmaxyIresult} are derived explicitly in Ref.~\cite{Supplmat} in terms of the microscopic parameters of the model. The anomalous Hall response, Eqs.~\eqref{eq:sigmaxyIresult}, constitutes the main result of our work. Unlike previous calculations of the AHE in multiband systems~\cite{KovalevSinova2009,KovalevSinova2010}, we included diagrams with crossed impurity lines as they equally contribute to the leading order approximation. The common physical origin of crossed diagrams manifests itself in the complementary contributions from the off-diagonal terms in $\underline x_{KK'}$ and $\underline \psi_{KK'}$. 

\paragraph{\textbf{Discussion.}} 
We now analyze our general result Eq.~\eqref{eq:sigmaxyIresult} in various simplifying cases. We consider both particular limits of the impurity potential $u(\v x)$, and special values of the parameters entering the clean Hamiltonian~\eqref{eq:H0}.

\paragraph{(i) Smooth disorder potential.} We first consider the case when $u(\v x)$ is smooth on the scale of the lattice constant $a$. Interband scattering is negligible and we obtain
\begin{equation}
\sigma_{xy} =  \sigma_{xy}^{(0)} (\vert \epsilon_\Gamma \vert/m_\Gamma) +  2 \sigma_{xy}^{(0)} (\vert \epsilon \vert/m). \label{eq:sigmaxySmoothDisorder}
\end{equation}
The anomalous Hall conductivity of a single Dirac cone is \cite{AdoTitov2015} 
\begin{equation}
\sigma_{xy}^{(0)} \left (\frac{\vert \epsilon \vert}{m}\right )= - \frac{e^2}{2h}\left [ \frac{16 \vert \epsilon \vert m^3 \theta(\epsilon^2 - m^2)}{(\epsilon^2 + 3m^2)^2} + \theta(m^2 - \epsilon^2)\right ]. \label{eq:sigmaxyADOT}
\end{equation}

\begin{figure}
\includegraphics[scale=.6]{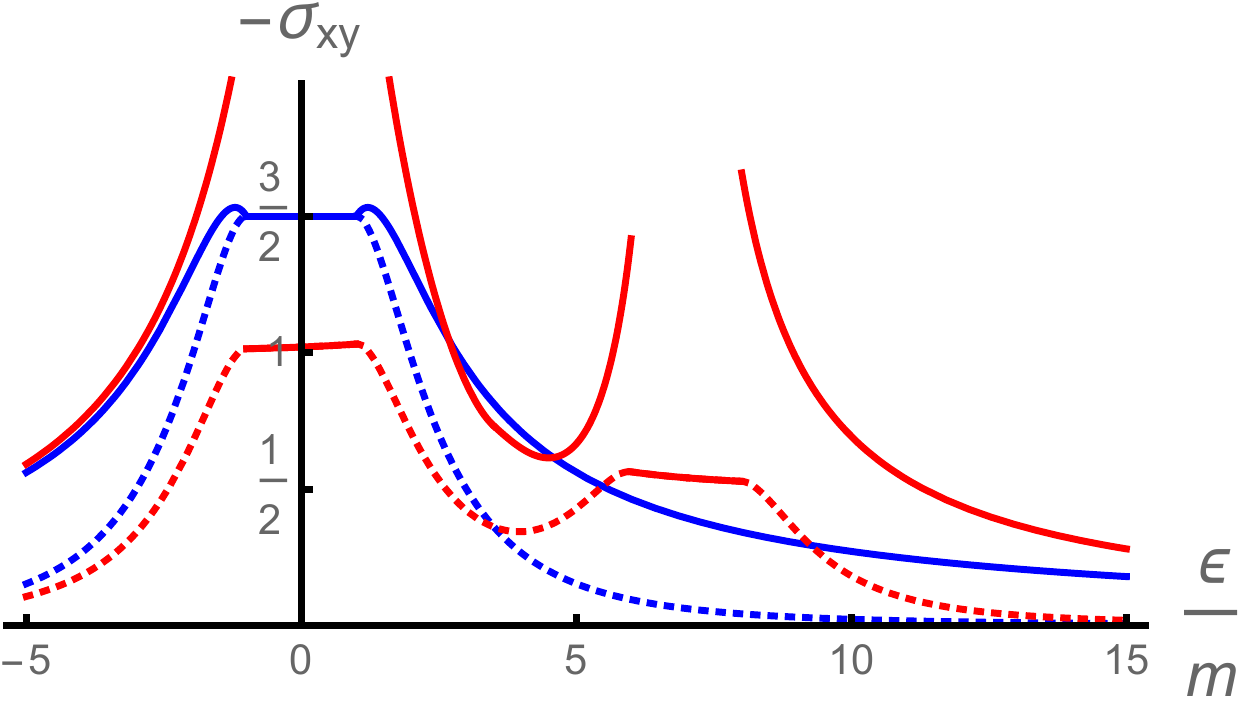} 
\caption{Comparison of the AHE in the cases of a smooth disorder potential (dotted) and point like scatterers (solid). We assumed $m_\Gamma = m>0$ and additionally imposed $v_\Gamma = v$ in the case of short ranged impurities. For the blue curves, we set $\epsilon_\Gamma = \epsilon$ while the red curves are obtained for $\epsilon_\Gamma = \epsilon - 7 m$.}
\label{fig:ConductancePlot}
\end{figure}

It should be noted that in this case there is no contribution from the $\Psi$ skew scattering diagrams [see Fig.~\ref{fig:Condbubbles} (d)] as they vanish. This pecularity is accidental and specific to the single Dirac cone limit. It can be traced back to the destructive interference of scattering from two-impurity complexes as evidenced from the plot of the probability $p_{AB}^\Psi$ in Fig.~\ref{fig:Amplitudes}. The result for smooth disorder potential is plotted in Fig.~\ref{fig:ConductancePlot} using dotted curves.

\paragraph{(ii) Fermi momentum in the gap.} In what follows we restore the possibility of nonzero interpocket scattering. When $\epsilon^2 < m^2$, i.e.~when the Fermi energy is in the gap of $X$ and $Y$ pockets, the problem essentially simplifies to a single Dirac cone and Eq.~\eqref{eq:sigmaxySmoothDisorder} holds again (using $ 2 \sigma_{xy}^{(0)} (\vert \epsilon \vert/m) = - e^2/h$).

Further, when $\epsilon_\Gamma^2 < m_\Gamma^2$ (Fermi energy in the gap of the $\Gamma$ pocket) the problem simplifies to two equal Dirac cones. Surprisingly, the resulting Hall conductivity is again given by Eq.~\eqref{eq:sigmaxySmoothDisorder} (using $  \sigma_{xy}^{(0)} (\vert \epsilon_\Gamma \vert/m_\Gamma) = - e^2/2h$) and is independent on the ratio $W_{XY}/W$.

\paragraph{(iii) Equal DPs.} We next consider the situation when the three DPs are equal, i.e.~$E_\Gamma = 0$, $v_\Gamma = v$ and $m_\Gamma = m$. The general expression for the Hall conductance is presented in Ref.~\cite{Supplmat}. We note that for smooth disorder potential, the effect of intraband scattering enters only to second order ($\epsilon^2 > m^2$)
\begin{equation}
\sigma_{xy} = \sigma_{xy}^{(0)} \left (\frac{\vert \epsilon \vert}{m}\right ) \left [3 +  \left (\frac{\epsilon^2}{m^2} -1\right ) \frac{W^2_{\Gamma X} + 2W_{\Gamma X} W_{XY}}{W^2} \right ]. \label{eq:sigmaxyEqualCones}
\end{equation}
While the single band Hall conductivity, Eq. \eqref{eq:sigmaxyADOT}, decays as $\vert \epsilon \vert ^{-3}$, for finite $W_{\Gamma X}$ we find a term which decays only as $\vert \epsilon \vert^{-1}$. Thus interband scattering strongly enhances anomalous Hall conductivity. This effect persists to the case of arbitrary interband scattering. The Hall conductance \eqref{eq:sigmaxyEqualCones} continuously approaches the gap value $\sigma_{xy} = -3 \sgn{m}e^2/2h$.

{\textit{(iv) Point like scatterers}}. When the impurity potential is short ranged on the scale of $a$, the matrix $\underline{W}_{K,K'} = n_{\rm imp} \vert u(0) \vert^2$ for all $K, K'$. The experimental analysis of the weak-antilocalization effect  \cite{DzeroGalitski2015} in SmB$_6$ samples  \cite{ThomasXia2013,NakajimaPaglione2013} suggests that this limit is most important for present day experiments.

The explicit formula of $\sigma_{xy} (\epsilon/m, \epsilon_\Gamma/m)$ for the case $v_\Gamma = v$, and $m_\Gamma = m$ has been relegated to Ref.~\cite{Supplmat}. 
If $\epsilon_\Gamma = \epsilon$ this result further simplifies to 
\begin{equation}
\sigma_{xy} = - \frac{e^2}{h}\frac{8 \vert\epsilon \vert m 
  \left( \epsilon ^2+8m^2\right)}{3 \left( \epsilon ^2+3m^2\right)^2}.
\end{equation}
A graphical comparison between the cases of smooth disorder and point like impurities is shown in Fig.~\ref{fig:ConductancePlot}. It should be noted, that in the limit of short range scatters our result for the Hall conductance ceases to be a continuous function: it displays discontinuities at $\epsilon_K = m_K$. A similar behavior was recently found in the Bychkov-Rashba model \cite{AdoTitov2016}. This is an artifact of an approximation that exploits the basic assumption $n_{\rm imp} \ll n_{\rm min}$. As a consequence, our result is not applicable in the energy window $\vert \epsilon_K - m_K\vert \lesssim \boldsymbol{\Gamma}$. A more elaborate calculation should reveal smoothening of the discontinuities in the immediate vicinity of the band edges. 

%%%%%%%%%%%%%%%%%%%%%%%%%%%%%%%%
%%%%%%%%%%%%%%%%%%%%%%%%%%%%%%%%
%%%%%%%%%%%%%%%%%%%%%%%%%%%%%%%%
%%%%%%%%%%%%%%%%%%%%%%%%%%%%%%%%

\paragraph{\textbf{Conclusion and outlook.}}
We have derived the anomalous Hall response Eq.~\eqref{eq:sigmaxyIresult} on the surface of a cubic topological Kondo insulator. We investigated a surface state model with three Dirac fermions plus an incipient forth $M$-band in the generic case allowing for unequal Fermi and Zeeman energies as well as unequal Fermi velocities. We have analyzed several limiting cases of our general result Eq.~\eqref{eq:sigmaxyIresult}. As a byproduct of this analysis, we found that a system of two equal Dirac cones (as it occurs in Graphene) displays an AHE which is universal and independent of details of the scattering potential.

Inasmuch experiments on TKIs are concerned, our most important conclusion is that the magnetization and gate voltage dependence of the AHE can be used to gain information about the microscopic nature of surface states and impurities. Indeed, the analysis of various limiting cases of the three-band Dirac model reveals that the large energy asymptote of the anomalous Hall response scales as $(m/\vert \epsilon \vert)^{3}$ in the case of smooth impurity potential while $\sigma_{xy} \sim m/\vert \epsilon \vert$ for short range scatterers. This behavior persists in the generic result. In contrast, in the Bychkov-Rashba model $\sigma_{xy} \sim m/\epsilon^{2}$ \cite{AdoTitov2016}.

As mentioned in the introduction, present day experimental samples are believed to host a multitude of large ferromagnetic domains. In our theory, smooth fluctuations of the magnetization can be taken into account by averaging the final result. Even after this procedure, the  asymptotics allow to distinguish smooth and sharp impurity potentials in the described manner. Up to now magnetotransport experiments on TKIs concentrated a hysteretic behavior in the longitudinal conductance. Systematic investigation of the transverse conductance is still needed.
 
In this paper we analyzed the semiclassical AHE and uncovered the importance of diffractive skew scattering in the context of topological Kondo insulators. In a parallel vein, our results have further rich consequences for anomalous transport phenomena in other multiband material systems such as Weyl semimetals \cite{Burkov} and chiral $p$-wave superconductors \cite{Lutchyn,Kallin,Andreev-Spivak}. Quantum effects, such as interaction and localization corrections to the conductivity tensor, the quantum AHE~\cite{WengFang2015,LiuQi2015,ChangLi2015} and the surface state quantum Hall effect~\cite{QiZhang2009,BruneMolenkamp2011,KoenigMirlin2014,YoshimiTokura2015} on TKIs remain a theoretical and experimental challenge for the future.

\paragraph{\textbf{Acknowledgements.}} 
We thank A. Andreev, I. Dmitriev, M.~Khodas, L.~Li, J.~Sauls, K.~Sun and M. Titov for important discussions. P.M.O. and E.J.K. acknowledge hospitality by the Department of Physics and Astronomy at Michigan State University,  and by the Department of Physics at University of Michigan (E.J.K.). This work was financially supported in part by NSF Grant No. DMR-1506547 (M.D.), and NSF Grants No. DMR-1606517 and ECCS-1560732 (E.J.K. and A.L.). Support for this research at the University of Wisconsin-Madison was provided by the Office of the Vice Chancellor for Research and Graduate Education with funding from the Wisconsin Alumni Research Foundation. Support for this research at Michigan State University was provided by the Institute for Mathematical and Theoretical Physics with funding from the office of the Vice President for Research and Graduate Studies.

%%%%%%%%%%%%%%%%%%%%%%%%%%%%%%%%%%%%%%
%%%%%%%%%%%%%%%%%%%%%%%%%%%%%%%%%%%%%%

%\bibliography{AHESmB20150925}

%%%%%%%%%%%%%%%%%%%%%%%%%%%%%%%%%%%%%%
%%%%%%%%%%%%%%%%%%%%%%%%%%%%%%%%%%%%%%

%%%%%%%%%% Merge with supplemental materials %%%%%%%%%%
\clearpage

\onecolumngrid

\begin{center}
\textbf{\large Supplementary Materials For \\  Anomalous Hall Effect on the surface of topological Kondo insulators}
\end{center}
%%%%%%%%%% Merge with supplemental materials %%%%%%%%%%
%%%%%%%%%% Prefix a "S" to all equations, figures, tables and reset the counter %%%%%%%%%%
\setcounter{equation}{0}
\setcounter{figure}{0}
\setcounter{table}{0}
\setcounter{page}{1}
\makeatletter
\renewcommand{\theequation}{S\arabic{equation}}
\renewcommand{\thefigure}{S\arabic{figure}}
\renewcommand{\bibnumfmt}[1]{[S#1]}
\renewcommand{\citenumfont}[1]{S#1}
%%%%%%%%%% Prefix a "S" to all equations, figures, tables and reset the counter %%%%%%%%%%

\section{Summary of results from Supplementary Material}

\subsection{Expressions entering Eq.~\eqref{eq:sigmaxyIresult} of the main text}

We here express all quantities of Eq.~\eqref{eq:sigmaxyIresult} of the main text in terms of microscopic parameters. The contributions from non-crossing diagrams are
\begin{subequations}
\begin{eqnarray}
\underline{F} &=& (1 - \underline{a} \underline{W})^{-1},\\
\underline{a} &=&  \sum_{K} \frac{(\epsilon_K^2 - m_K^2)\pi \nu_K(\epsilon_K)}{4 \epsilon_K \left (\epsilon_K \boldsymbol \Gamma_K + m_K \boldsymbol \Gamma_K^{(m)}\right )}\underline{\Pi}_K, \label{eq:a}  \\
\underline{b} &=& -\sum_{K} \frac{ (\boldsymbol \Gamma_K m_K + \epsilon_K \boldsymbol \Gamma_K^{(m)})\pi \nu_K(\epsilon_K)}{2 \epsilon_K \left (\epsilon_K \boldsymbol \Gamma_K + m_K \boldsymbol \Gamma_K^{(m)}\right )} \underline{\Pi}_K  , \label{eq:b}
\end{eqnarray}
while diagrams (c) and (d) of Fig.~\ref{fig:Condbubbles} involve
\begin{eqnarray}
\underline{x}_{KK} &=&  \sum_{K'} \frac{\underline{W}_{KK'}^2}{v_Kv_K'} c_{(K,K')} d_{(K,K')} , \\
\underline{\psi}_{KK} &=& 2 \sum_{K'}  \frac{\underline{W}_{KK} \underline{W}_{KK'}}{v_Kv_K'} c_{(K,K')} \tilde d_{(K,K')} f_{(K,K')} , \\
\underline{x}_{KK'}&\stackrel{K \neq K'}{=}&  \frac{\theta(\epsilon_K\epsilon_K')}{v_K^2}  [c_{(K,K')} \tilde d_{(K,K')} (1- f_{(K',K)}) ] \underline W_{KK}\underline W_{KK'} + K \leftrightarrow K' \label{eq:xOffDiago},\\
\underline{\psi}_{KK'} &\stackrel{K \neq K'}{=}& \frac{\theta(-\epsilon_K\epsilon_K')}{v_K^2}  [c_{(K,K')} \tilde d_{(K,K')} (1- f_{(K',K)}) ] \underline W_{KK}\underline W_{KK'} + K \leftrightarrow K' \label{eq:psiOffDiago}.
\end{eqnarray}
\label{eq:WholeAHEresponse}
\end{subequations}
Note the complementary contributions of $\underline{x}_{KK'}$ and $\underline{\psi}_{KK'}$. We introduced the following dimensionless functions
\begin{subequations}
\begin{eqnarray}
c_{(K,K')} &=& 2 \pi \nu_{K'}(\epsilon_{K'}) \, m_{K'}/p_{F,K}^2,\\
d_{(K,K')} &=&  [{\epsilon_K}/{\epsilon_{K'}} + {m_K}/{m_{K'}} ], \\
\tilde d_{(K,K')} &=&   [{\epsilon_K}/{\epsilon_{K'}} + {v_K}/{v_{K'}} ] ,\\
f_{(K,K')} &=& \theta(p_{F,K} - p_{F,K'}) [1 - p_{F,K'}^2/p_{F,K}^2].
\end{eqnarray}
\label{eq:WholeAHEAuxiliary}
\end{subequations}

\subsection{Anomalous Hall effect for equal Dirac pockets}

In this section we present the general formula for the AHE in the case of equal Dirac pockets, i.e. $E_\Gamma = 0$, $v_\Gamma = v$ and $m_\Gamma = m$:

\begin{eqnarray}
\sigma_{xy} (\vert \epsilon \vert /m) &=& 3 \sigma_{xy}^{(0)} (\vert \epsilon \vert /m) + \frac{4 A b}{\left(A \left(2 W_{\Gamma X}^2-W^2-W W_{XY}\right)+\left(W^2+3 W W_{\Gamma X}+W W_{XY}+2 W_{\Gamma X}^2+2 W_{\Gamma X} W_{XY}\right)\right)^2} \notag \\
&& \times \Big [4 W W_{\Gamma X} W_{XY} (W+W_{XY})+  W_{\Gamma X}^2 \left(2 W^2-\frac{\left(4 A^2+A-2\right) W_{XY}^2}{(A-1)^2}-\frac{12 (A+1) W W_{XY}}{A-1}\right) \notag \\
&&+\frac{W_{\Gamma X}^3  \left(2 (A (4 A+7)+10) W_{XY}-4 \left(A^2+A-2\right) W\right)}{(1-A)^2}+\frac{(A (2 A+11)+2) W_{\Gamma X}^4}{(1-A)^2}\Big ]
\end{eqnarray}
Here we introduced the notation
\begin{subequations}
\begin{eqnarray}
A &=& \frac{\epsilon^2 - m^2}{2 ( \epsilon^2 + m^2)}, \\
b &=& - \frac{\vert \epsilon \vert m}{2 ( \epsilon^2 + m^2)} .
\end{eqnarray}
\end{subequations}

These expressions are the origin of Eq.~\eqref{eq:sigmaxyEqualCones} of the main text.

\subsection{Anomalous Hall response in the limit of point-like scatterers}

{We here present the formula for the anomalous Hall response in the case of equal velocities $v_\Gamma = v$ and equal Zeeman field $m_\Gamma = m$ in $\Gamma$, $X$ and $Y$ pockets. Then the Hall conductivty is a function of two parameters $\bar \epsilon = \epsilon/m$ and $\bar \epsilon_\Gamma = \epsilon_\Gamma/m$, only. Outside of any gap we obtain
\begin{equation}\label{eq:sigmaxyEqualScatt}
\sigma_{xy}\! =\! \sgn{\epsilon}\! \left[ \theta(\epsilon^2 - \epsilon_\Gamma^2) \sigma_{xy}^{[1, \sgn{\epsilon\epsilon_\Gamma}]}\! + \theta(\epsilon_\Gamma^2 - \epsilon^2) \sigma_{xy}^{[2, \sgn{\epsilon\epsilon_\Gamma}]}\right] 
\end{equation}
with 
%\begin{widetext}
\begin{subequations}
\begin{align}
\sigma_{xy}^{[1,+]} &=- 4\frac{e^2}{h} [ 8 \bar \epsilon ^6 \bar \epsilon_\Gamma +4 \bar \epsilon ^5 \left(5 \bar \epsilon_\Gamma ^2+4\right)+2 \bar \epsilon ^4 \bar \epsilon_\Gamma  \left(7 \bar \epsilon_\Gamma ^2+38\right)+\bar \epsilon ^3 \left(-\bar \epsilon_\Gamma ^4+242 \bar \epsilon_\Gamma ^2+63\right)+2 \bar \epsilon ^2 \bar \epsilon_\Gamma  \left(2 \bar \epsilon_\Gamma ^4+82 \bar \epsilon_\Gamma ^2+261\right) \notag \\
&+\bar \epsilon  \left(7 \bar \epsilon_\Gamma ^6+38 \bar \epsilon_\Gamma ^4+315 \bar \epsilon_\Gamma ^2+288\right)+2 \bar \epsilon_\Gamma  \left(\bar \epsilon_\Gamma ^6+2 \bar \epsilon_\Gamma ^4+9 \bar \epsilon_\Gamma ^2+72\right) ] / [4 \bar \epsilon ^3 \bar \epsilon_\Gamma +4 \bar \epsilon ^2 \left(\bar \epsilon_\Gamma ^2+2\right)+\bar \epsilon  \bar \epsilon_\Gamma  \left(\bar \epsilon_\Gamma ^2+23\right)+5 \bar \epsilon_\Gamma ^2+27]^2  , \\
\sigma_{xy}^{[1,-]}&= - 4\frac{e^2}{h} [ -8 \bar \epsilon ^6 \bar \epsilon_\Gamma -4 \bar \epsilon ^5 \bar \epsilon_\Gamma ^2-2 \bar \epsilon ^4 \bar \epsilon_\Gamma  \left(7 \bar \epsilon_\Gamma ^2+10\right)+\bar \epsilon ^3 \left(33 \bar \epsilon_\Gamma ^4+26 \bar \epsilon_\Gamma ^2-3\right)-2 \bar \epsilon ^2 \bar \epsilon_\Gamma  \left(5 \bar \epsilon_\Gamma ^4+12 \bar \epsilon_\Gamma ^2-16\right) \notag \\
&-5 \bar \epsilon  \left(\bar \epsilon_\Gamma ^2-2\right) \left(\bar \epsilon_\Gamma ^2-1\right)^2+2 \bar \epsilon_\Gamma  \left(\bar \epsilon_\Gamma ^2-1\right)^3] / {\left[\bar \epsilon_\Gamma  \left(\bar \epsilon  \left((\bar \epsilon_\Gamma -2 \bar \epsilon )^2+7\right)-3 \bar \epsilon_\Gamma \right)+3\right]^2}, \\
\sigma_{xy}^{[2,+]} &= - 4\frac{e^2}{h} [8 \bar \epsilon ^7+8 \bar \epsilon ^6 \bar \epsilon_\Gamma -2 \bar \epsilon ^5 \left(\bar \epsilon_\Gamma ^2-8\right)+4 \bar \epsilon ^4 \bar \epsilon_\Gamma  \left(3 \bar \epsilon_\Gamma ^2+19\right)+\bar \epsilon ^3 \left(19 \bar \epsilon_\Gamma ^4+242 \bar \epsilon_\Gamma ^2+63\right)+2 \bar \epsilon ^2 \bar \epsilon_\Gamma  \left(4 \bar \epsilon_\Gamma ^4+82 \bar \epsilon_\Gamma ^2+261\right) \notag \\
&+\bar \epsilon  \left(\bar \epsilon_\Gamma ^6+38 \bar \epsilon_\Gamma ^4+315 \bar \epsilon_\Gamma ^2+288\right)+2 \bar \epsilon_\Gamma  \left(2 \bar \epsilon_\Gamma ^4+9 \bar \epsilon_\Gamma ^2+72\right)]/\left[4 \bar \epsilon ^3 \bar \epsilon_\Gamma +4 \bar \epsilon ^2 \left(\bar \epsilon_\Gamma ^2+2\right)+\bar \epsilon  \bar \epsilon_\Gamma  \left(\bar \epsilon_\Gamma ^2+23\right)+5 \bar \epsilon_\Gamma ^2+27\right]^2, \\
\sigma_{xy}^{[2,-]} &=- 4\frac{e^2}{h} [8 \bar \epsilon ^7-8 \bar \epsilon ^6 \bar \epsilon_\Gamma +\bar \epsilon ^5 \left(8-10 \bar \epsilon_\Gamma ^2\right)-4 \bar \epsilon ^4 \bar \epsilon_\Gamma  \left(3 \bar \epsilon_\Gamma ^2+2\right)+\bar \epsilon ^3 \left(25 \bar \epsilon_\Gamma ^4+18 \bar \epsilon_\Gamma ^2+1\right)-2 \bar \epsilon ^2 \bar \epsilon_\Gamma  \left(5 \bar \epsilon_\Gamma ^2 \left(\bar \epsilon_\Gamma ^2+4\right)-18\right)\notag \\
&+\bar \epsilon  \left(\bar \epsilon_\Gamma ^6+20 \bar \epsilon_\Gamma ^4-29 \bar \epsilon_\Gamma ^2+10\right)-2 \left(\bar \epsilon_\Gamma ^5-\bar \epsilon_\Gamma ^3+\bar \epsilon_\Gamma \right)]/\left[\bar \epsilon_\Gamma  \left(\bar \epsilon  \left((\bar \epsilon_\Gamma -2 \bar \epsilon )^2+7\right)-3 \bar \epsilon_\Gamma \right)+3\right]^2.
\end{align}
%\end{widetext}
\end{subequations}

%\section*{Appendix}
\section{Supplementary Material: Calculation of $\sigma_{xy}^I$}

In this supplementary material we present details regarding the calculation of Eq.~\eqref{eq:sigmaxyIresult} and ~\eqref{eq:WholeAHEresponse} of the main text.

\subsection{Average Green's function}
We first present the calculation of the average Green's function in the limit $n_{\rm imp.} \ll n_{\rm min}$. In this limit the retarded self-energy approximately becomes
\begin{equation}
\hat \Sigma^R (\epsilon) \simeq \sum_{\v R_j}\hat t^R_{\v R_j} (\epsilon).
\end{equation}
Here, $\lbrace \v R_j \rbrace$ are the impurity positions and the T-matrix $t^R_{\v R_j}$ of a single impurity is calculated at the level of Born approximation, see Fig.~\ref{fig:Born}. Upon disorder average denoted by angular brackets we obtain

\begin{eqnarray}
\left \langle [\underline t_{\v R_j}^R (\epsilon)]_{\v p, \v p'} \right \rangle &=& \left \langle \underline u_{\v R_j} (\v p - \v p') \right \rangle  + \left  \langle \int_{\v p''} \underline u_{\v R_j} (\v p - \v p'') \underline G^R_0( \v p'', \epsilon ) \underline u_{\v R_j} (\v p'' - \v p')\right \rangle \notag \\
&=&  \sum_{{K,K'}}  \int_{\v p''}u(\v p + \v Q_{K}- \v p'' - \v Q_{K'}) G_{K',0}^R( \v p'', \epsilon) u(\v p'' + \v Q_{K'} - \v p' - \v Q_{K} )   \frac{\underline \Pi_{K} }{V} (2\pi)^2 \delta(\v p - \v p') 
\end{eqnarray}
We introduced the symbol $\int_\v p = \int d^2p/(2\pi)^2$.
Momentum conservation (reobtained after impurity average) forbids off-diagonal matrix elements in Dirac pocket (DP) space. We absorb the real part of the self-energy into a redefinition of $E_K$ and $m_K$. The imaginary part of the self-energy leads to the scattering rates presented in Eqs.~\eqref{eq:Gammas} of the main text.

In our calculations, we formally do not only include the Born approximation diagram, Fig.~\ref{fig:Born}, but also the resummation of rainbow diagrams (self consistent Born approximation) and diagrams with a single intersection of impurity lines. To the leading order, this does not alter the result for the scattering rates, Eqs.~\eqref{eq:Gammas}.

%%%%%%%%%%%%%%%%%%%%%%%
%%%%%%%%%%%%%%%%%%%%%%%
\begin{figure}[b]
\includegraphics[scale=1]{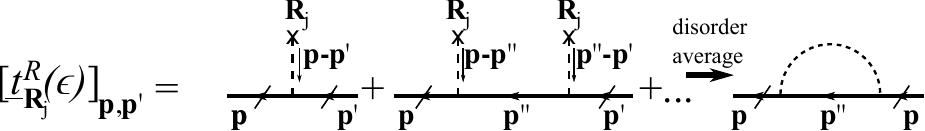} 
\caption{T-matrix of a single impurity at the level of Born approximation, before and after disorder average.}
\label{fig:Born}
\end{figure}
%%%%%%%%%%%%%%%%%%%%%%%
%%%%%%%%%%%%%%%%%%%%%%%

\subsection{Current vertex and non-crossing approximation}

We first consider the dressed velocity vertices. In view of momentum conservation it is useful to represent the left (right) current vertices as row (column) vectors in DP space, and, at the same time, as vertices in spin space, i.e.~in formulas
\begin{equation}
\hat j^{\mathcal L}_\mu = \underline{\v j} \sigma_\mu \text{ and } \hat j^{\mathcal R}_\mu = \underline{\v j} ^T \sigma_\mu 
\end{equation}
where, on the bare level, $\underline {\v j} \rightarrow \underline{\v v} = (v_\Gamma, v, v)$.

For the resummation of impurity potentials at the left vertex, we will need the following integral (which is a matrix in spin space and a matrix element of the diagonal matrices $\underline I^{\mathcal {L,R}}_\mu$ in DP space)
\begin{equation}
[\underline I^{\mathcal L}_{\mu}]_{KK} := \int_{\v p} G_K^A(\v p, \epsilon) \sigma_\mu G_K^R(\v p, \epsilon) \simeq \sigma_\mu [\underline a]_{KK} + i \sigma_\mu \sigma_z [\underline b]_{KK}.
\end{equation}
Similarly, we also need for the right vertex the object
\begin{equation}
[\underline I^{\mathcal R}_{\mu}]_{KK} := \int_{\v p} G_K^R(\v p, \epsilon) \sigma_\mu G_K^A(\v p, \epsilon) \simeq \sigma_\mu [\underline a]_{KK} + i \sigma_z \sigma_\mu [\underline b]_{KK}.
\end{equation}
The diagonal matrices $\underline{a}$ and $\underline{b}$ are presented in Eq.~\eqref{eq:a} and \eqref{eq:b} of the main text.
We readily find
\begin{equation}
\hat j_\mu^{\mathcal L} \simeq  \underline{\v v} \underline F \sigma_\mu + i \sigma_\mu \sigma_z \underline{\v v} \underline F \underline b \underline W \underline F
\end{equation}
and analogously 
\begin{equation}
\hat j_\mu^{\mathcal R} \simeq ( \underline{\v v} \underline F )^T\sigma_\mu + i \sigma_z \sigma_\mu (\underline{\v  v} \underline F \underline b \underline W \underline F)^T.
\end{equation}
The result for $\sigma_{xy}^I$ in the non-crossing approximation immediately follows 
\begin{equation}
[\sigma_{xy} ]_{nc} = \frac{e^2}{h} \tr^{\sigma}[ \hat j_x^{\mathcal L} \underline I_y^{\mathcal R} \underline{\v v}^T] = \frac{e^2}{h} 2 \underline{\v v} \underline F \underline b \underline F^T \underline{\v v}^T.
\end{equation}

\subsection{X and $\Psi$ diagrams}

We will distinguish diagonal (``intraband'') and off-diagonal (``interband'')parts of the matrices $\underline X = 2 \underline a \underline x \underline a$  and $\underline \Psi = 2 \underline a \underline \psi \underline a$ introduced in Eq.~\eqref{eq:sigmaxyIresult} of the main text. For the evaluation of X and $\Psi$ diagrams we define the following quantities :
\begin{equation}
N_K(\v p) = \epsilon_K + m_K \sigma_z + v_K \v p \cdot \boldsymbol \sigma.
\end{equation}
and (assuming $\epsilon_K^2 > m_K^2$)
\begin{subequations}
\begin{eqnarray}
\mathcal{G}_K^{R/A}(\v p) &=& \frac{1}{(\epsilon_K \pm i 0)^2 - (v_K p)^2 - m_K^2}, \\
\mathcal{G}_K^{R/A}(\v r) & =& \int_{\v p} e^{i \v p \v r} \mathcal{G}_K^{R/A}(\v p) = \frac{1}{4 v_K^2} \left [Y_0(p_{F,K}r) \mp i \sgn{\epsilon_K} J_0(p_{F,K}r)\right ],
\end{eqnarray}
\end{subequations}
Here, $J_0(x)$ and $Y_0(x)$ are the zeroth Bessel functions of first and second kind. 

In the evaluation of diagrams, we use $\sigma_{xy} = -\sigma_{yx}$ and the fact that only the symmetric part of $3 \times 3$ matrices $\underline{X}$ and $\underline \Psi$ enters the Hall conductance. This section of the supplementary material will be devoted to the calculation of those matrices. The matrices $\underline{X}$ and $\underline \Psi$ are diagrammatically represented by diagrams analogous to Fig.~\ref{fig:Condbubbles}, (b)-(d), of the main text, but the vertex correction should be omitted and is incorporated separately in the final result.

\subsubsection{Momentum conservation}

%%%%%%%%%%%%%%%%%%%%%%%%	
%%%%%%%%%%%%%%%%%%%%%%%%
	\begin{figure}
	\begin{center}
	\includegraphics[scale=.6]{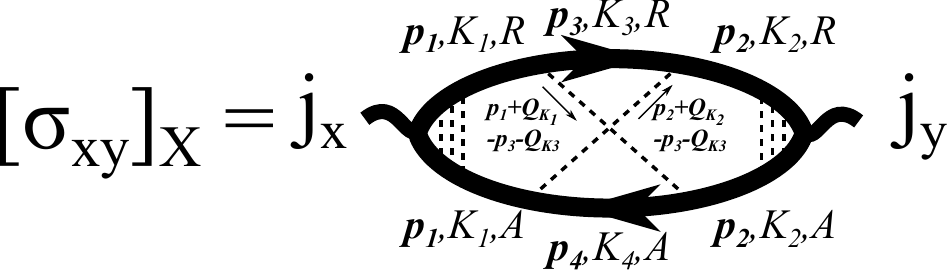} \quad	\includegraphics[scale=.6]{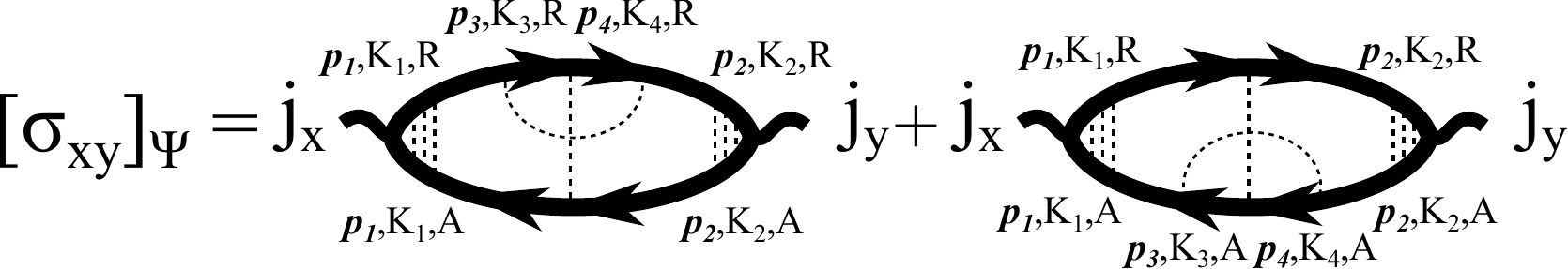} 
	\caption{The X and $\Psi$- diagrams labelled with momentum variables, pocket index $K$ and retarded/advanced (R/A) labels. Momentum conservation is analyzed in Eqs.~\eqref{eq:MomentaXdiag} and \eqref{eq:MomentaPsidiag}.}
	\label{fig:XandPsidiagram}
	\end{center}
	\end{figure}
%%%%%%%%%%%%%%%%%%%%%%%%
%%%%%%%%%%%%%%%%%%%%%%%%

We label X and $\Psi$-diagrams with momentum variables, see Fig.~\ref{fig:XandPsidiagram}. We readily conclude from momentum conservation that for the X diagram
	\begin{equation}
	\v p_1 + \v Q_{K_1} + 	\v p_2 + \v Q_{K_2} = 	\v p_3 + \v Q_{K_3} + 	\v p_4 + \v Q_{K_4}.
	\end{equation}
	Since we assume distant Fermi surfaces, $p_{F,K} \ll \pi/a$, this implies
	\begin{subequations}
		\begin{eqnarray}
	 \v Q_{K_1}  + \v Q_{K_2} &=& \v Q_{K_3}  + \v Q_{K_4} , \label{eq:MomentaXdiagQ} \\
		\v p_1  + 	\v p_2  &=& \v p_3  + \v p_4 . 
	\end{eqnarray}
	\label{eq:MomentaXdiag}
	\end{subequations}

		Analogously we analyze the $\Psi$ diagram. As compared to the X diagram, momentum conservation and the assumption of distant Fermi surfaces now imply
		\begin{subequations}
		\begin{eqnarray}
	 \v Q_{K_1} - \v Q_{K_2} &=& \v Q_{K_3}  - \v Q_{K_4} , \label{eq:MomentaPsidiagQ} \\
		\v p_1  -	\v p_2  &=& \v p_3  - \v p_4 . 
	\end{eqnarray}
	\label{eq:MomentaPsidiag}
	\end{subequations}
	Note that Eqs.~\eqref{eq:MomentaXdiagQ} and \eqref{eq:MomentaPsidiagQ} are to be understood modulo a reciprocal lattice vector (Umklapp scattering). In particular, there is an Umklapp $\Psi$ diagram in which $K_1 = K_4 \neq K_2 = K_3$.

	\subsubsection{Integrals over Bessel functions}

The evaluation of X and $\Psi$ diagrams involves various integrals over Bessel functions. These are 
\begin{subequations}
\begin{eqnarray}
I_A^{(1)}(r) &=& \int_0^\infty ds J_1^2(rs)J_1(s)Y_0(s) = \frac{1}{2\pi}\left [\theta (1-r) +\frac{\theta(r-1)}{r^2}\right ] , \\
I_B^{(1)}(r) &=& \int_0^\infty ds J_1^2(rs)J_0(s)Y_1(s) = - 1/\pi +  I_A^{(1)}(r) ,\label{eq:BesselIB1} \\
I_C^{(1)}(r) &=& \int_0^\infty ds J_1(s)J_1(rs)J_0(s)Y_1(rs) = -I_A^{(1)}(r), \\
I_B^{(2)}(r) &=& \int_0^\infty ds J_1'(s) J_1(s) Y_0 (rs) J_0(rs) = \frac{1}{2} \left [I_A^{(1)}(1/r)+I_B^{(1)}(1/r)\right ] . \label{eq:BesselIB2} 
\end{eqnarray}
\end{subequations}
Details on the evaluation of such integrals can be found in Ref.~\cite{AdoTitov2016}.

\subsubsection{X diagram: Intraband contribution}

For this contribution we find
\begin{eqnarray}
		 \underline{X}_{KK} &=& \sum_{K'} W_{KK'}^2 \int_{\lbrace \v p_i \rbrace} (2\pi)^2 \delta( \v p_1 + \v p_2 - \v p_3 - \v p_4) \frac{i \mathfrak I \mathcal G^R_K(\v p_1) \, \mathfrak I \mathcal G^R_K(\v p_2)\, \mathfrak I \mathcal G^R_{K'}(\v p_3)\,\mathfrak R \mathcal G^R_{K'}(\v p_4) }{[2(\epsilon_K \boldsymbol \Gamma_K + m_K \boldsymbol \Gamma^{(m)}_K)]^2} \notag \\
		 && \times \tr\Big [N_K(\v p_1)\sigma_x N_K(\v p_1) N_{K'}(\v p_3)  N_K(\v p_2)\sigma_y N_K(\v p_2) N_{K'}(\v p_4) \notag \\
		 && \phantom{\tr \Big [}- N_K(\v p_1)\sigma_x N_K(\v p_1) N_{K'}(\v p_4)  N_K(\v p_2)\sigma_y N_K(\v p_2) N_{K'}(\v p_3)  \Big ] \notag \\
		 &\doteq& \sum_{K'} W_{KK'}^2 \int_{\lbrace \v p_i \rbrace} (2\pi)^2 \delta( \v p_1 + \v p_2 - \v p_3 - \v p_4) \frac{i \mathfrak I \mathcal G^R_K(\v p_1) \, \mathfrak I \mathcal G^R_K(\v p_2)\, \mathfrak I \mathcal G^R_{K'}(\v p_3)\,\mathfrak R \mathcal G^R_{K'}(\v p_4) }{[2(\epsilon_K \boldsymbol \Gamma_K + m_K \boldsymbol \Gamma^{(m)}_K)]^2} \notag \\
		 && \times 16 i v_K^3 v_{K'}\left (m_{K'} \epsilon_K + m_{K} \epsilon_{K'} \right )
		 p_{1,x} p_{2,y} [\v p_1-\v p_2]\wedge[\v p_3- \v p_4]\notag \\
		 &=&\sum_{K'} W_{KK'}^2 2\pi \sgn{\epsilon_{K'}} \frac{ (m_{K'} \epsilon_K + m_{K} \epsilon_{K'}  )  (\epsilon_K^2 - m_K^2) }{(4 v_K v_{K'})^3 [\epsilon_K \boldsymbol \Gamma_K + m_K \boldsymbol \Gamma^{(m)}_K]^2} [I_A^{(1)}(p_{F,K}/p_{F,K'}) - I_B^{(1)}(p_{F,K}/p_{F,K'}) ] .%\\ &=& \sum_{K'} 2 \underline{a}_{KK} x_{KK'K} \underline{a}_{KK}.
		\end{eqnarray}

Here, and in the evaluation of all other X and $\Psi$ diagrams, the symbol $\doteq$ denotes that momenta $\v p_1 $ and $\v p_2$ are to be taken on-shell, such that $(v_K p_{1,2})^2 = \epsilon_K^2 - m_K^2$. The notation $\v a \wedge \v b = \epsilon_{\mu \nu} a_\mu b_\nu$ is used. We readily see from the first line, that contributions arise only if both $\epsilon_K^2 > m_K^2$ and $\epsilon_{K'}^2 >m_{K'}^2$: if the Fermi energy is in the gap of a certain pocket, then this pocket does not contribute to $\underline{X}_{KK}$. The same is true for all other X and $\Psi$ diagrams, too. 

\subsubsection{X diagram: Interband contribution}

Let's now consider the interband contribution from X diagram:

\begin{eqnarray}
	\left . \underline{X}_{KK'} \right \vert_{K \neq K'} &=&   W_{KK}W_{KK'} \int_{\lbrace \v p_i \rbrace} (2\pi)^2 \delta( \v p_1 + \v p_2 - \v p_3 - \v p_4) \frac{\epsilon_{\mu \nu}}{2} \notag \\
	&& \frac{i \mathfrak I \mathcal G^R_K(\v p_1) \, \mathfrak I \mathcal G^R_{K'}(\v p_2)\,\left ( \mathfrak I \mathcal G^R_{K}(\v p_3)\,\mathfrak R \mathcal G^R_{K'}(\v p_4) - \mathfrak R \mathcal G^R_{K}(\v p_3)\,\mathfrak I \mathcal G^R_{K'}(\v p_4) \right  ) }{[2(\epsilon_{K'} \boldsymbol \Gamma_{K'} + m_{K'} \boldsymbol \Gamma^{(m)}_{K'})][2(\epsilon_K \boldsymbol \Gamma_K + m_K \boldsymbol \Gamma^{(m)}_K)]} \notag \\
		 && \times \tr\Big [N_K(\v p_1)\sigma_\mu N_K(\v p_1) N_{K}(\v p_3)  N_{K'}(\v p_2)\sigma_\nu N_{K'}(\v p_2) N_{K'}(\v p_4) \notag \\
		 && \phantom{\tr \Big [}- N_K(\v p_1)\sigma_\mu N_K(\v p_1) N_{K'}(\v p_4)  N_{K'}(\v p_2)\sigma_\nu N_{K'}(\v p_2) N_{K}(\v p_3)  \Big ] \notag \\
		 &\doteq &  W_{KK}W_{KK'} \int_{\lbrace \v p_i \rbrace} (2\pi)^2 \delta( \v p_1 + \v p_2 - \v p_3 - \v p_4) \frac{\epsilon_{\mu \nu}}{2} \notag \\
	&& \frac{i \mathfrak I \mathcal G^R_K(\v p_1) \, \mathfrak I \mathcal G^R_{K'}(\v p_2)\,\left ( \mathfrak I \mathcal G^R_{K}(\v p_3)\,\mathfrak R \mathcal G^R_{K'}(\v p_4) - \mathfrak R \mathcal G^R_{K}(\v p_3)\,\mathfrak I \mathcal G^R_{K'}(\v p_4) \right  ) }{[2(\epsilon_{K'} \boldsymbol \Gamma_{K'} + m_{K'} \boldsymbol \Gamma^{(m)}_{K'})][2(\epsilon_K \boldsymbol \Gamma_K + m_K \boldsymbol \Gamma^{(m)}_K)]} \notag \\
	&& \times 16 i v_K^2 \v p_1 \wedge \v p_2 \Big [ 2 m_K \epsilon_K v_{K'}^2 \v p_2 \wedge \v p_4  +  2 m_{K'} \epsilon_{K'} v_{K}^2 \v p_1 \wedge \v p_3 \notag \\
	&& +v_Kv_{K'} \left (\lbrace  \v p_3 \wedge \v p_2 + \v p_4 \wedge \v p_1 \rbrace \lbrace m_K \epsilon_{K'} + m_{K'} \epsilon_K\rbrace + \lbrace  \v p_2 \wedge \v p_1 + \v p_3 \wedge \v p_4 \rbrace \lbrace m_K \epsilon_{K'} - m_{K'} \epsilon_K\rbrace\right )  \Big ] \notag \\
	&=& W_{KK} W_{KK'} \frac{2 \pi }{\left (4 v_K v_{K'}\right )^3[\epsilon_K \boldsymbol \Gamma_K + m_K \boldsymbol \Gamma^{(m)}_K][\epsilon_{K'} \boldsymbol \Gamma_{K'} + m_{K'} \boldsymbol \Gamma^{(m)}_{K'}]} \notag \\
	&& \times \left \lbrace m_{K'} \left (\epsilon_{K'} + \frac{v_K'}{v_K} \epsilon_K \right ) \left (\epsilon_K^2 - m_K^2\right ) \left [\sgn{\epsilon_{K'}} I_A^{(1)}\left (\frac{p_{F,K}}{p_{F,K'}}\right )- \sgn{\epsilon_{K}} I_C^{(1)}\left (\frac{p_{F,K}}{p_{F,K'}}\right )\right ]\right \rbrace\notag \\
	&& + K \leftrightarrow K' . \label{eq:Supp:Xinter}
		\end{eqnarray}

\subsubsection{$\Psi$ diagram: Intraband contribution}

This contribution is
\begin{eqnarray}
\underline \Psi_{KK}  &=&\sum_{K'} \underline W_{KK}  \underline W_{KK'} \int_{\lbrace \v p_i \rbrace}(2\pi)^2 \delta( \v p_1 - \v p_2 - \v p_3 + \v p_4) \frac{i \mathfrak I \mathcal G^R_K(\v p_1) \, \mathfrak I \mathcal G^R_K(\v p_2)\, \mathfrak I \mathcal G^R_{K'}(\v p_3)\,\mathfrak R \mathcal G^R_{K'}(\v p_4) }{[2(\epsilon_K \boldsymbol \Gamma_K + m_K \boldsymbol \Gamma^{(m)}_K)]^2} \notag \\ 
&&\times \epsilon_{\mu \nu} \tr\Big [N_K(\v p_1)\sigma_\mu N_K(\v p_1) N_{K'}(\v p_3) N_{K'}(\v p_4) N_K(\v p_2)\sigma_\nu N_K(\v p_2)  \notag \\ 
&& \phantom{\tr \Big [ \epsilon_{\mu \nu}}+ N_K(\v p_1)\sigma_\mu N_K(\v p_1) N_{K'}(\v p_4) N_{K'}(\v p_3) N_K(\v p_2)\sigma_\nu N_K(\v p_2)   \Big ] \notag \\ 
 &\doteq&\sum_{K'} \underline W_{KK}  \underline W_{KK'} \int_{\lbrace \v p_i \rbrace}(2\pi)^2 \delta( \v p_1 - \v p_2 - \v p_3 + \v p_4) \frac{i \mathfrak I \mathcal G^R_K(\v p_1) \, \mathfrak I \mathcal G^R_K(\v p_2)\, \mathfrak I \mathcal G^R_{K'}(\v p_3)\,\mathfrak R \mathcal G^R_{K'}(\v p_4) }{[2(\epsilon_K \boldsymbol \Gamma_K + m_K \boldsymbol \Gamma^{(m)}_K)]^2} \notag \\ 
 && \times 16i v_K^2 \v p_1 \wedge \v p_2 \Big [2v_{K'}^2  m_K \epsilon_K \v p_3 \wedge \v p_4  - 2v_{K}^2  m_{K'} \epsilon_{K'} \v p_1 \wedge \v p_2 \notag \\
&&+v_Kv_{K'} \left (\lbrace  \v p_1 \wedge \v p_3 + \v p_4 \wedge \v p_2 \rbrace \lbrace m_K \epsilon_{K'} + m_{K'} \epsilon_K\rbrace + \lbrace  \v p_1 \wedge \v p_4 + \v p_3 \wedge \v p_2 \rbrace \lbrace m_K \epsilon_{K'} - m_{K'} \epsilon_K\rbrace\right )  \Big ] \notag \\
&=& \sum_{K'} \underline W_{KK}  \underline W_{KK'} (- 4\pi) \sgn{\epsilon_{K'}} \frac{\epsilon_K^2 - m_K^2}{\left (4 v_K v_{K'}\right )^3[\epsilon_K \boldsymbol \Gamma_K + m_K \boldsymbol \Gamma^{(m)}_K]^2} \notag \\
&& \times \left (2 m_{K'} \epsilon_{K'} \frac{v_K}{v_{K'}} I_{B}^{(2)}\left (\frac{p_{F,K'}}{p_{F,K}}\right ) + \epsilon_K m_{K'}\left [I_A^{(1)}\left (\frac{p_{F,K}}{p_{F,K'}}\right ) + I_B^{(1)}\left (\frac{p_{F,K}}{p_{F,K'}}\right )\right ]\right ).
\end{eqnarray}

Note the similarity between this expression and $\underline X_{KK'}\vert_{K \neq K'}$.

\subsubsection{$\Psi$ diagram: Straight interband contribution}

We first consider the case of straight scattering in the interband contribution, i.e.~$K_1 = K_3 \neq K_2 = K_4$. Antisymmetrization of $\sigma_{xy}$ and symmetrization of $\underline{\Psi}_{KK'}$ lead to

\begin{eqnarray}
	\left . \underline{\Psi}_{KK'} \right \vert_{\substack{K \neq K'\\\text{straight}}} &=&   W_{KK}W_{KK'} \int_{\lbrace \v p_i \rbrace} (2\pi)^2 \delta( \v p_1 - \v p_2 - \v p_3 + \v p_4) \frac{\epsilon_{\mu \nu}}{2}  \notag \\
	&& \frac{i \mathfrak I \mathcal G^R_K(\v p_1) \, \mathfrak I \mathcal G^R_{K'}(\v p_2)\,\left ( \mathfrak I \mathcal G^R_{K}(\v p_3)\,\mathfrak R \mathcal G^R_{K'}(\v p_4) + \mathfrak R \mathcal G^R_{K}(\v p_3)\,\mathfrak I \mathcal G^R_{K'}(\v p_4) \right  ) }{ [2(\epsilon_{K'} \boldsymbol \Gamma_{K'} + m_{K'} \boldsymbol \Gamma^{(m)}_{K'})][2(\epsilon_K \boldsymbol \Gamma_K + m_K \boldsymbol \Gamma^{(m)}_K)]} \notag \\
		 && \times \tr\Big [N_K(\v p_1)\sigma_\mu N_K(\v p_1) N_{K}(\v p_3)  N_{K'}(\v p_4) N_{K'}(\v p_2)\sigma_\nu N_{K'}(\v p_2) \notag \\
		 && \phantom{\tr \Big [}- N_K(\v p_1)\sigma_\mu N_K(\v p_1) N_{K'}(\v p_2)\sigma_\nu N_{K'}(\v p_2)  N_{K'}(\v p_4) N_{K}(\v p_3)  \Big ] \notag \\
		 &\doteq & W_{KK}W_{KK'} \int_{\lbrace \v p_i \rbrace} (2\pi)^2 \delta( \v p_1 - \v p_2 - \v p_3 + \v p_4) \notag \\
	&& \frac{i \mathfrak I \mathcal G^R_K(\v p_1) \, \mathfrak I \mathcal G^R_{K'}(\v p_2)\,\left ( \mathfrak I \mathcal G^R_{K}(\v p_3)\,\mathfrak R \mathcal G^R_{K'}(\v p_4) + \mathfrak R \mathcal G^R_{K}(\v p_3)\,\mathfrak I \mathcal G^R_{K'}(\v p_4) \right  ) }{2[2(\epsilon_{K'} \boldsymbol \Gamma_{K'} + m_{K'} \boldsymbol \Gamma^{(m)}_{K'})][2(\epsilon_K \boldsymbol \Gamma_K + m_K \boldsymbol \Gamma^{(m)}_K)]} \notag \\
	&& \times 16 i v_K v_{K'} \v p_1 \wedge \v p_2\Big [ 2m_K \epsilon_K v_{K'}^2 \v p_4 \wedge \v p_2 + 2 m_{K'} \epsilon_{K'} v_K^2 \v p_1 \wedge \v p_3 \notag \\
&& -v_Kv_{K'} \left (\lbrace  \v p_1 \wedge \v p_2 + \v p_4 \wedge \v p_3 \rbrace \lbrace m_K \epsilon_{K'} + m_{K'} \epsilon_K\rbrace + \lbrace  \v p_2 \wedge \v p_3 + \v p_1 \wedge \v p_4 \rbrace \lbrace m_K \epsilon_{K'} - m_{K'} \epsilon_K\rbrace\right )  \Big ] \notag \\
& = & W_{KK} W_{KK'} \frac{-2 \pi }{\left (4 v_K v_{K'}\right )^3[\epsilon_K \boldsymbol \Gamma_K + m_K \boldsymbol \Gamma^{(m)}_K][\epsilon_{K'} \boldsymbol \Gamma_{K'} + m_{K'} \boldsymbol \Gamma^{(m)}_{K'}]} \notag \\
&& \times \left \lbrace m_{K'} \left (\epsilon_{K'} + \frac{v_K'}{v_K} \epsilon_K \right ) \left (\epsilon_K^2 - m_K^2\right ) \left [\sgn{\epsilon_{K'}} I_A^{(1)}\left (\frac{p_{F,K}}{p_{F,K'}}\right )+ \sgn{\epsilon_{K}} I_C^{(1)}\left (\frac{p_{F,K}}{p_{F,K'}}\right )\right ]\right \rbrace\notag \\
	&& + K \leftrightarrow K' .
		\end{eqnarray}

Note the cancellation between this diagram and $X_{KK'}\vert_{K \neq K'}$.

\subsubsection{$\Psi$ diagram: Umklapp interband contribution}

We now turn our attention to the interband Umklapp process, again we symmetrize $\underline{\Psi}$ in pocket space and use antisymmetry of $\sigma_{xy}$

\begin{eqnarray}
	\left . \underline{\Psi}_{KK'} \right \vert_{\substack{K \neq K'\\\text{Unklapp}}} &=&   (W_{KK'})^2 \int_{\lbrace \v p_i \rbrace} (2\pi)^2 \delta( \v p_1 - \v p_2 - \v p_3 + \v p_4) \frac{\epsilon_{\mu \nu}}{2}  \notag \\
	&& \frac{i \mathfrak I \mathcal G^R_K(\v p_1) \, \mathfrak I \mathcal G^R_{K'}(\v p_2)\,\left ( \mathfrak I \mathcal G^R_{K'}(\v p_3)\,\mathfrak R \mathcal G^R_{K}(\v p_4) + \mathfrak R \mathcal G^R_{K'}(\v p_3)\,\mathfrak I \mathcal G^R_{K}(\v p_4) \right  ) }{ [2(\epsilon_{K'} \boldsymbol \Gamma_{K'} + m_{K'} \boldsymbol \Gamma^{(m)}_{K'})][2(\epsilon_K \boldsymbol \Gamma_K + m_K \boldsymbol \Gamma^{(m)}_K)]} \notag \\
		 && \times \tr\Big [N_K(\v p_1)\sigma_\mu N_K(\v p_1) N_{K'}(\v p_3)  N_{K}(\v p_4) N_{K'}(\v p_2)\sigma_\nu N_{K'}(\v p_2) \notag \\
		 && \phantom{\tr \Big [}- N_K(\v p_1)\sigma_\mu N_K(\v p_1) N_{K'}(\v p_2)\sigma_\nu N_{K'}(\v p_2)  N_{K}(\v p_4) N_{K'}(\v p_3)  \Big ] \notag \\
		 &=&   (W_{KK'})^2 \int_{\lbrace \v p_i \rbrace} (2\pi)^2 \delta( \v p_1 - \v p_2 - \v p_3 + \v p_4)   \notag \\
	&& \frac{i \mathfrak I \mathcal G^R_K(\v p_1) \, \mathfrak I \mathcal G^R_{K'}(\v p_2)\,\left ( \mathfrak I \mathcal G^R_{K'}(\v p_3)\,\mathfrak R \mathcal G^R_{K}(\v p_4) + \mathfrak R \mathcal G^R_{K'}(\v p_3)\,\mathfrak I \mathcal G^R_{K}(\v p_4) \right  ) }{2 [2(\epsilon_{K'} \boldsymbol \Gamma_{K'} + m_{K'} \boldsymbol \Gamma^{(m)}_{K'})][2(\epsilon_K \boldsymbol \Gamma_K + m_K \boldsymbol \Gamma^{(m)}_K)]} \notag \\
	&& \times(- 16 i)v_K^2v_{K'}^2 \v p_1 \wedge \v p_2 [\v p_2 - \v p_3] \wedge [\v p_1 - \v p_4] \left (m_K \epsilon_{K'} + m_{K'} \epsilon_K \right ) \notag \\
	&=& (W_{KK'})^2 \frac{-2 \pi (\epsilon_K m_{K'} + m_K \epsilon_{K'})}{\left (4 v_K v_{K'}\right )^3[\epsilon_K \boldsymbol \Gamma_K + m_K \boldsymbol \Gamma^{(m)}_K][\epsilon_{K'} \boldsymbol \Gamma_{K'} + m_{K'} \boldsymbol \Gamma^{(m)}_{K'}]} \notag \\
&& \times  \frac{v_K'}{v_K} \left (\epsilon_K^2 - m_K^2\right ) \left [\sgn{\epsilon_{K'}} I_A^{(1)}\left (\frac{p_{F,K}}{p_{F,K'}}\right )+ \sgn{\epsilon_{K}} I_C^{(1)}\left (\frac{p_{F,K}}{p_{F,K'}}\right )\right ]\notag \\
&& + K \leftrightarrow K' \notag \\
&=&0. \label{eq:Supp:PsiUmklapp}
		\end{eqnarray}

\section{Contributions to $\sigma_{xy}^I$ from the  $M$ point}
\label{eq:MPoint}

Since the calculation of $\sigma_{xy}$ in the Kubo formalism always involves one Green's function which is off-shell, on needs in principle the knowledge of the single particle spectrum in the entire Brilloin zone and thus a more general Hamiltonian than Eq.~\eqref{eq:H0} of the main text. 

Indeed, for an off-shell state around the $\tilde K'$ point entering $X_{KK}$ or $\Psi_{KK}$ diagrams we need to know the spectrum for all quasimomenta $k \leq 2p_{F,K} + p_{F,K'}$ around the $K'$ point. The same bound appears in the analysis of interband $X$ and $\Psi$ diagrams.

One can also analyze diagrams where the off-shell contribution stems from states with the radius $p_{F,\Gamma} + 2 p_{F} $ around the $M$ point of the Brillouin zone. At the M-point the spectrum is gapped, as we will argue next, this is the reason why such diagrams can be neglected. 

For concreteness we will assume a Green's function of $M$-point states of the form $[\Delta + \v p^2/2m_M]^{-1}$ with $\Delta>0$ the distance in energy space to the closest states. The integrand of Bessel functions for diagrams involving states near the $M$-point is smaller as compared to the case of only $\Gamma$, $X$ and $Y$ pockets by a factor
\begin{equation}
\frac{2 m_M v_{F,K}^2}{ \lbrace \epsilon_K, m_K, v_{F,K} p_{F,K} \rbrace} \left \vert \frac{K_\nu \left (\frac{\sqrt{2 m_M \Delta}}{p_{F,K} } s\right )}{Y_\nu (s)} \right \vert .
\end{equation}
To determine an $s$-independent parameter, we use that $X$ and $\Psi$ diagrams are dominated by $s \sim 1$.
To summarize, contributions from the $M$-point are small as long as one of the following conditions holds:
\begin{subequations}
\begin{equation}
v_{F,K}^2 m_M \ll \lbrace \vert m_K \vert,\sqrt{\epsilon_K^2 - m_K^2}\rbrace \quad \text{(algebraic suppression)},
\end{equation}
or 
\begin{equation}
\frac{\sqrt{2 m_M \Delta}}{p_{F,K}} \gg 1 \quad \text{(exponential suppression)}.
\end{equation}
 \label{eq:unimportanceMpoint}
\end{subequations}
In conclusion, we require either a sufficiently large gap at the M-point or considerable flatness of the spectrum. 

In conclusion, this supplementary material was devoted to the derivation of the anomalous Hall response presented in Eqs.~\eqref{eq:sigmaxyIresult} of the main text and \eqref{eq:WholeAHEresponse} of this supplementary material. There, the contribution of diagrams with crossed impurities, Eqs.~\eqref{eq:Supp:Xinter}-\eqref{eq:Supp:PsiUmklapp} of this section, was rewritten with the use of Eqs.~\eqref{eq:WholeAHEAuxiliary}.

\section{Fermion number fractionalization and $\sigma_{xy}^{II}$}

We here present the connection between $\sigma_{xy}^{II}$ and fermion number fractionalization. We concentrate on the case when the Fermi energy is inside the magnetization induced gap. As we explained in the main text, disorder is unimportant and it is thus sufficient to restrict ourselves to a single Dirac cone. We consider a domain wall between inverted and trival band structure, and here disregard details on the origin of the band inversion. 

We first omit the presence of a finite surface magnetization (it will be restored later) and describe the 3D TI by

\begin{equation}
H_{M(z)} = \sum_{\mu = x,y}[\v p - e\v A(x,y)]_\mu  \sigma_\mu \sigma_z \tau_y + v p_z  \sigma_z \tau_x + M(z) \tau_z. \label{eq:HamKink}
\end{equation}

Both $\sigma_\mu$ (spin index) and $\tau_\mu$ (parity index) are Pauli matrices, $e$ is the electron charge and we set the speed of light $c = 1$ here and in the following. The Hamiltonian anticommutes with $\sigma_z \tau_y$ and thus the spectrum is invariant with respect to reflection about zero energy $E=0$. The mass $M(z) \stackrel{\vert z \vert \rightarrow \infty}{\longrightarrow} \text{sign}(z)\, M_\infty$ mimics the band inversion ($M_\infty >0$). 
In the 1D case, i.e. omitting $x,y$ coordinates, there must be at least one zero mode per spin orientation localized near the kink (Callias-Bott-Seeley theorem). In the 3D case with a magnetic field $B>0$ perpendicular to the interface, the Hamiltonian is block diagonal in the space of 2D Dirac Landau levels.
In this case, the number of zero modes must be at least $g = BA\vert e\vert /(2\pi)$ (the degeneracy of the zeroth Landau level, $A$ is the area penetrated by the flux). Normalizability of the wave function implies that the zero mode 4-spinors contain only 2 independent entries $u_{n,0} (x,y,z) = (\psi_n(x,y), i \sigma_z \psi_n(x,y))^T$ ($n = 1, \dots, g$).

We follow the standard arguments to further show the fermion number fractionalization and compare the topological case of a kink ($M(z)$ as indicated above) and the trivial case without any kink ($M(z) = \text{const.}$). 
The completeness of eigenstates, $\lbrace \Psi^{top.}_E, u_{n,0} \rbrace$ and $\lbrace \Psi^{triv.}_E \rbrace$, respectively, implies

\begin{eqnarray}
0 &=& \left (\sum \hspace*{-.5cm} \int_{-\infty}^{0^-} dE \vert \Psi^{top.}_E ( \v x )\vert ^2 + \sum_n \vert u_{n,0}( \v x )\vert^2 + \sum \hspace*{-.5cm} \int_{0^+}^{+\infty} dE \vert  \Psi^{top.}_E (\v x) \vert ^2 \right )- \sum \hspace*{-.5cm} \int_{-\infty}^{+\infty} dE \vert  \Psi^{triv.}_E (\v x) \vert ^2 \notag \\
&\stackrel{\lbrace H, \sigma_z \tau_y\rbrace = 0}{=}& 2 \left [\sum \hspace*{-.5cm} \int_{-\infty}^{0^-} dE \left (\vert \Psi^{top.}_E ( \v x )\vert ^2 - \vert \Psi^{triv.}_E (\v x) \vert ^2\right ) + \sum_n\frac{\vert u_{n,0}( \v x )\vert^2}{2}\right ].
\end{eqnarray}

Then one can calculate the relative fermion number comparing topological and non-topological situation:
\begin{equation}
N \equiv  \int d^dx \sum \hspace*{-.5cm} \int_{-\infty}^{0^-} dE \left (\vert \Psi^{top.}_E ( \v x )\vert ^2 - \vert \Psi^{triv.}_E (\v x) \vert ^2\right ) =  -\frac{1}{2} \sum_n \int d^dx\vert u_{n,0}( \v x )\vert^2  =  \frac{BA  e}{4\pi}.
\end{equation}

Thus, using Eq.~\eqref{eq:KuboStreda} of the main text, we find for massless surface Dirac fermions in a magnetic field with chemical potential just below (above) zero $\sigma_{xy}^{II} = e^2/2h$ ($\sigma_{xy}^{II} = -e^2/2h$). 

Let's now return to the situation, when a small, but finite magnetization is present ($0<m \ll M_\infty$)
\begin{equation}
H = H_{M(z)} + m \tau_y. \label{eq:KinkWithMagn}
\end{equation}
Of course, for $m \neq 0$ the chiral symmetry is broken, the Callias-Bott-Seeley theorom is inapplicable and the surface Dirac electrons are gapped. Indeed, in the absence of magnetic fields, the low energy theory of Eq.~\eqref{eq:KinkWithMagn} is described by the Hamiltonian  $H_{E \ll M_\infty} = v \v p \cdot \boldsymbol{ \sigma} + m \sigma_z$ (cf. Eq.~\eqref{eq:H0} of the main text). 

For the calculation of $\sigma_{xy}^{II}$, recall that $m >0$ leads to a little down-shift of the zeroth Landau level. Therefore, as in the case of infinitesimally positive chemical potential, we find $\sigma_{xy}^{II} = -e^2/2h$. This result persists in the limit $B \rightarrow 0$ and concludes the derivation of Eq.~\eqref{eq:sigmaxygap} of the main text.


\begin{thebibliography}{38}
\bibitem{FootnoteInterference}
Maximally crossed diagrams can form an exception, since they lead to weak localization corrections that are enhanced logarithmically.
\end{thebibliography}
\end{document}